\documentclass[cameraready]{Interspeech}
\usepackage{times}
\usepackage{latexsym}
\usepackage[T1]{fontenc}

\usepackage[utf8]{inputenc}

\usepackage{microtype}

\usepackage{inconsolata}

\usepackage{graphicx}

\usepackage{enumitem}
\usepackage{amsmath}
\usepackage{tabularx}
\usepackage{pifont}
\usepackage{tcolorbox}
\usepackage{booktabs}
\usepackage{multirow}
\usepackage{multicol}
\usepackage{cite}
\let\citep\cite
\let\citet\cite
\usepackage{comment}

\makeatletter

\newcommand{\Rmnum}[1]{\expandafter\@slowromancap\romannumeral #1@}
\makeatother

\title{Audio Language Model for Deepfake Detection Grounded in Acoustic Chain-of-Thought
}

\author[equalcontribution]{Runkun}{Chen}
\author[equalcontribution]{Yixiong}{Fang}
\author[equalcontribution]{Pengyu}{Chang}
\author[equalcontribution]{Yuante}{Li}
\author{Massa}{Baali}
\author{Bhiksha}{Raj}

\address{
    Carnegie Mellon University, USA
}

\email{runkunc@andrew.cmu.edu}


\begin{document}

\maketitle

\begin{abstract}
Deepfake speech detection systems are often limited to binary classification tasks and struggle to generate interpretable reasoning or provide context-rich explanations for their decisions. These models primarily extract latent embeddings for authenticity detection but fail to leverage structured acoustic evidence such as prosodic, spectral, and physiological attributes in a meaningful manner. This paper introduces \textsc{CoLMbo-DF}, a Feature-Guided Audio Language Model that addresses these limitations by integrating robust deepfake detection with explicit acoustic chain-of-thought reasoning. By injecting structured textual representations of low-level acoustic features directly into the model prompt, our approach grounds the model's reasoning in interpretable evidence and improves detection accuracy. To support this framework, we introduce a novel dataset of audio pairs paired with chain-of-thought annotations. Experiments show that our method, trained on a lightweight open-source language model, significantly outperforms existing audio language model baselines despite its smaller scale, marking a significant advancement in explainable deepfake speech detection.
\end{abstract}
\section{Introduction}

Audio deepfake detection is a critical challenge in speech forensics, 
with applications in security, fraud prevention, misinformation 
detection, and speaker verification~\citep{yi2023audio}.
Traditional approaches have relied on signal-level 
artifacts, such as inconsistencies in pitch~\citep{de2012synthetic, 
pal2018synthetic}, formant trajectories~\citep{sahidullah2015comparison}, 
energy patterns~\citep{dimitriadis2005auditory}, and voice quality 
statistics~\citep{javed2022voice}. These methods, however, are inflexible; 
their performance heavily depend on feature engineering, and may degrade under domain shift. More recent works have explored applying deep learning to audio deepfake detection~\citep{li2025surveydeepfake, aasist, tak2022automatic, guo2024audio}. While these methods exhibit strong performance on existing data, they offer limited interpretability, which could be problematic in high-stakes forensic settings.

Recent advances in multimodal modeling have explored the use of Audio 
Language Models (ALM), but their application to deepfake detection remains 
limited. Existing Speaker Language Models (SLMs), such as 
CoLMbo~\citep{baali2025colmbo} and SpeakerLM~\citep{yin2025speakerlmendtoendversatilespeaker}, 
primarily focus on speaker profiling. 
These models typically make decisions in a single step, without 
explicitly articulating how acoustic evidence supports the final 
judgment, limiting their utility for high-stakes forensic settings 
where human validation and accountability are required.

A key requirement in audio deepfake detection is grounding model 
decisions in measurable acoustic evidence. Purely implicit embeddings 
are high-dimensional and abstract, whereas reliable forensic reasoning 
demands structured, semantically grounded observations over prosodic, 
spectral, temporal, and voice quality cues. This grounding is critical 
for reliable detection but is not adequately addressed by existing ALMs, 
which often fail to generalize to unseen synthesizers or provide 
transparent decision rationales.

To address these challenges, we present \textsc{CoLMbo-DF}, a 
lightweight Feature-Guided Audio Language Model designed for audio deepfake detection. By injecting structured acoustic evidence directly into the 
language model input, our approach trains the model to produce 
feature-grounded chain-of-thought reasoning before emitting its final 
decision. New acoustic feature types can be incorporated simply by 
augmenting the training data, without modifying the model architecture.

\textsc{CoLMbo-DF} employs a projection-based architecture, where 
audio embeddings from a pretrained encoder are mapped into the language 
model's input space. Structured textual descriptions of low-level 
acoustic features are injected alongside the audio prefix to ground the 
model's reasoning in interpretable evidence. To support this formulation, 
we curate a new dataset \textsc{FakeReason} which consists of audio pairs with chain-of-thought annotations, 
built from ASVspoof 2019~\citep{asvproof} and VoxCeleb2~\citep{chung2018voxceleb2}, 
augmented with recent TTS systems. Experiments show that feature-guided 
reasoning improves robustness under synthesizer shift while requiring 
only a small base language model.

\textsc{CoLMbo-DF} outperforms existing pretrained audio 
language model baselines on ASVspoof 2019 despite its smaller scale, 
demonstrating that better acoustic grounding outperforms larger parameter 
counts. In summary, our key contributions are as follows:

\begin{itemize}[leftmargin=1.2em]
    \item \textbf{We propose \textsc{CoLMbo-DF}, a lightweight 
    feature-guided ALLM pipeline for Audio Deepfake Detection} that integrates 
    structured acoustic reasoning into an audio language model, 
    surpassing existing pretrained ALM baselines on ASVspoof 2019 
    with a smaller base model.
    
    \item \textbf{We curate a new dataset \textsc{FakeReason} of audio pairs and 
    chain-of-thought annotations} using ASVspoof 2019, VoxCeleb2, 
    state-of-the-art TTS models, and a text-only LLM for CoT generation.

\end{itemize}
\section{CoLMbo-DF Architecture}

\begin{figure*}[ht]
    \centering
    \includegraphics[width=0.7\linewidth]{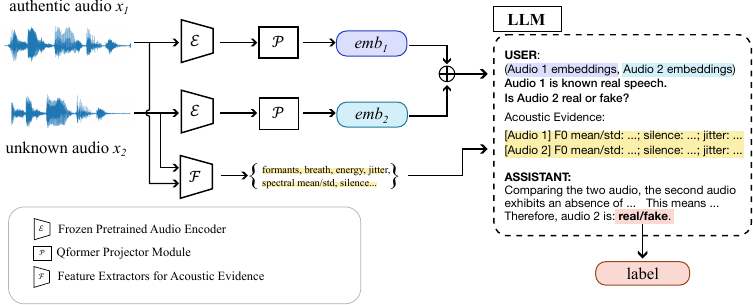}
    \caption{An overview of our proposed pipeline. Given an input pair, the audios are encoded and then projected by the pretrained audio encoder $\mathcal{E}$ and the projector module $\mathcal{P}$. Separately, acoustic features are extracted and exported to textual form as acoustic evidence. These are combined into the input of LLM, which uses reasoning to generate the final prediction.}
    \label{fig:pipeline}
\end{figure*}

\subsection{Architecture overview}
\label{subsec:arch}

Figure~\ref{fig:pipeline} presents our reasoning-centric audio-LLM pipeline.
The key design is to decouple acoustic representation learning from decision reasoning:
a strong pretrained speech encoder provides compact acoustic evidence, while an LLM performs structured reasoning
conditioned on both latent embeddings and explicit, interpretable acoustic attributes.

Our architecture contains three modules:
(i) a pretrained audio encoder $\mathcal{E}$, 
(ii) a lightweight trainable projector $\mathcal{P}$ that aligns audio representations to the LLM embedding space,
and (iii) an instruction-tuned LLM as the reasoning core.
We use pretrained WavLM-plus~\cite{wavlm} model as the audio encoder. Given an utterance waveform $x$, the encoder produces frame-level representations
$\mathbf{H}=\mathcal E(x)\in\mathbb{R}^{T\times d_e}$, which are pooled into a single vector
$\mathbf{h}=\mathrm{Pool}(\mathbf{H})\in\mathbb{R}^{d_e}$.
The projector maps $\mathbf{h}$ to a latent embedding prefix compatible with the LLM input space:
\begin{equation*}
\label{eq:projector}
\mathbf{Z}=\mathcal P(\mathbf{h})\in\mathbb{R}^{m\times d_\ell},
\end{equation*}
where $m$ is the number of projected tokens. We prepend $\mathbf{Z}$ to the textual prompt so that the LLM
can attend to audio evidence through standard self-attention, without modifying the LLM architecture. 
During inference, the model is prompted to generate outputs in a structured format to facilitate the extraction of the predicted label.

\subsection{Acoustic Evidence-grounded Reasoning}
In addition to latent embeddings, we also extract interpretable acoustic features to assist the model reasoning. In audio forensics, analysts may use defined audio features, such as pitch and formants, to identify synthesized speech.~\cite{li2025surveydeepfake} Inspired by this, we have provided the LLM with a selection of audio features for grounded reasoning. 

The features are serialized into a structured text block and included in the input as explicit acoustic evidence.
This encourages the LLM to ground its rationale in measurable attributes rather than relying solely on opaque embeddings.
The explicit evidence also acts as a mechanism for feature disentanglement, forcing the model to separately consider independent variables, such as pitch and pauses, before making a decision. 




We train the model via supervised fine-tuning (SFT) with next token prediction.
Each training instance consists of an input audio pair $(x_{1}, x_{2})$ and their extracted acoustic features, which are serialized and concatenated with the projected audio embeddings. The model is expected to output a chain-of-thought representing its reasoning process, followed by a structured output that includes the predicted label (real or fake).

\section{\textsc{FakeReason} Dataset}
To train our model, we constructed the \textsc{FakeReason} dataset, consisting of audio pairs and chain-of-thought annotations regarding the authenticity of the audio. 
Each audio pair consists of one authentic speech as reference, and the other as the subject of classification. 





\begin{figure}[h]                                                  
  \centering                                                       
  \includegraphics[width=\linewidth]{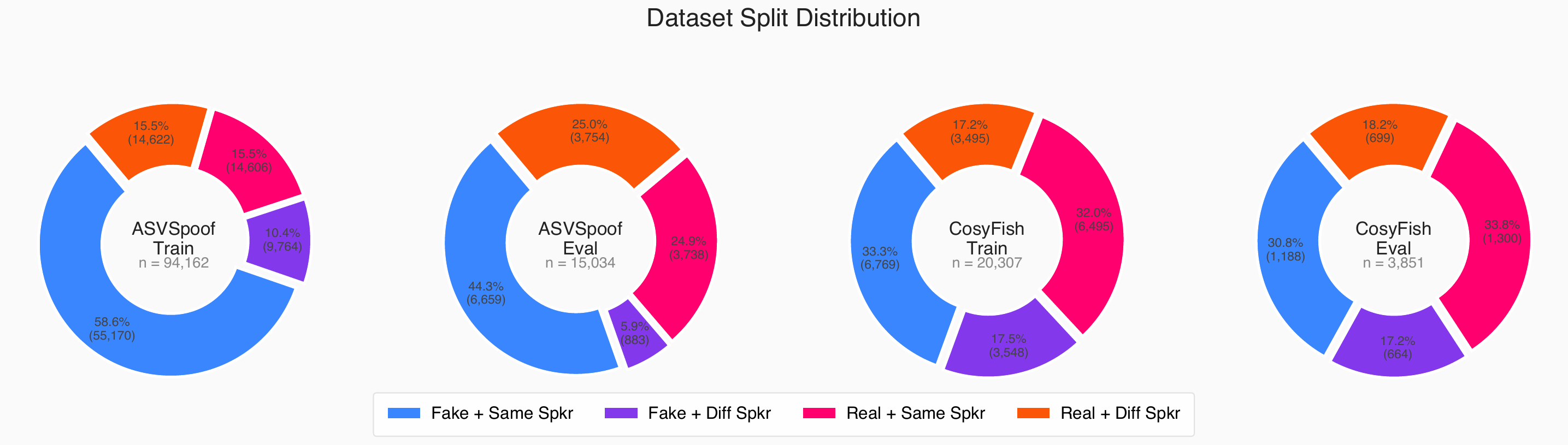}                                                                          
  \caption{Dataset Splits and Sample Counts}                       
  \label{fig:datasetsplit}                                        
\end{figure}

For the audio pairs in our dataset, we use the Logical Access (LA) partition of ASVspoof 2019~\cite{asvproof},
and expanded the training data by moving the original evaluation split into training and using the original development split for evaluation.
The final statistics are reported in Figure~\ref{fig:datasetsplit} (ASVspoof partition).


To reflect recent advances in speech synthesis, we generate additional spoofed audio using two state-of-the-art TTS models, Fish-Speech~\cite{liao2024fish} and CosyVoice2~\cite{du2024cosyvoice}. Using reference speech and transcripts sampled from VoxCeleb2~\cite{chung2018voxceleb2}, we synthesize $\sim$20K utterances per model and apply filtering to obtain $\sim$10K high-quality deepfakes for training. 
We sample additional utterance pairs from VoxCeleb2 as bonafide data. Dataset statistics are summarized in Figure~\ref{fig:datasetsplit} under the \texttt{CosyFish} partition.


To construct the chain-of-thought annotations in our dataset, we used a more capable LLM (Qwen3-30B-A3B-Instruct-2507~\cite{yang2025qwen3technicalreport}) to generate reasonings based on the acoustic evidence and the ground-truth label from samples in ASVspoof 2019 and \texttt{CosyFish}; the model was instructed to produce the reasoning \emph{as if} the ground truth were unavailable. We then manually edited the outputs to remove any explicit references to this assumption. A sample of the prompt can be found in Figure~\ref{fig:prompt}.
At inference time, we use the same prompt template but omit the ground-truth label.

\begin{figure}
    \centering
    \begin{tcolorbox}[colback=gray!5,colframe=gray!40]
    \tiny
    You are an expert in audio forensics and deepfake detection. \\
    Analyze the following pair of audio samples based on their acoustic features. \\
    The first audio is the reference audio you are comparing against. The second audio is an unknown sample. 
    Use the provided ground truth to keep your reasoning consistent with reality. \\
    \textbf{Audio 1}: \\
    Label: \texttt{(Genuine or Deepfake)} \\
    Features: \texttt{(Formatted features)} \\
    \textbf{Audio 2}: \\
    Label: \texttt{(Genuine or Deepfake)} \\
    Features: \texttt{(Formatted features)} \\
    \textbf{Ground Truth (for training reference, DO NOT contradict)}: \\
    - Audio 1 authenticity: \texttt{(Genuine or Deepfake)} \\
    - Audio 2 authenticity: \texttt{(Genuine or Deepfake)} \\
    - Speaker relationship: \texttt{(Same speaker or Different Speakers)} 

    \textbf{Task}: Produce a single, coherent reasoning narrative (no numbered steps) that covers: \\
    - Key acoustic traits of Audio 1 (pitch, formants, voice quality, prosody) \\
    - Key acoustic traits of Audio 2 \\
    Important similarities and differences between the two recordings \\
    -  Assessment of the second audio’s authenticity (genuine vs. deepfake) with evidence \\
    - Justification of whether the speakers are the same person or different people \\
    - An explicit acknowledgment that the reasoning aligns with the provided ground truth
    
    Keep the reasoning faithful to the ground truth, while explaining how acoustic evidence supports it. \\
    You MUST PRETEND LIKE you DO NOT KNOW the ground truth labels when analyzing the features, but your final conclusion MUST MATCH the ground truth. 
    
    \textbf{Final Conclusion (use this exact structure at the end):} \\
    - Speaker 1: [Genuine / Deepfake] \\
    - Speaker 2: [Genuine / Deepfake] \\
    - Speaker Relationship: [Same Speaker / Different Speakers] \\
    - Reasoning: [A shorter rephrasing of your reasoning process 
                  in natural language, around 100 words] 
    
    \end{tcolorbox}
    \caption{Prompt Template for generating synthesized CoT.}
    \label{fig:prompt}
\end{figure}

\section{Experimental Setup}

As a baseline, we use WavLM-base-plus~\cite{wavlm}\footnote{\url{https://huggingface.co/microsoft/wavlm-base-plus}} as the audio encoder, a 6-layer QFormer network~\cite{qformer} as the projector,
and Llama~3.2-1B-Instruct~\cite{llama32,llama3models} as the LLM. Unless stated otherwise, the LLM is kept frozen during training by default.

We evaluated our model on accuracy and F1 score. It should be noted that due to the nature of LLM generation, which predicts discrete token \textit{sequences} as outputs, we are unable to use metrics like EER that require non-discrete probabilities.

We study three prompting variants for reasoning supervision: 1) \textbf{CoT}, using the full chain-of-thought texts from the \textsc{FakeReason} dataset; 2) \textbf{ShortCoT}, using only a summary of the chain-of-thought instead of full text; and 3) \textbf{NoCoT}, which trains the model on only structured output for label prediction, without any reasoning. We additionally evaluate an \textbf{Unfreeze} setting that fine-tunes the full model. For ablation studies, we also incorporate a label for speaker verification into the structured output during training and inference. Finally, we tested training on different subsets of the \textsc{FakeReason} dataset by excluding CosyFish from the training data in some of our experiments.

\section{Results and Analysis}\label{subsec:results}

We report our experiment results in comparison with baselines in Table \ref{tab:results}. We chose Qwen2-Audio-Instruct ~\cite{qwenaudio} as our baseline: Qwen2-Audio-Instruct is an instruction-finetuned LLM trained on general audio-related tasks (but \textit{not} containing deepfake detection). For the evaluation, we used Accuracy and F1 score as our metrics, for reasons explained in the previous section. 

\begin{table}[h]
    \centering
    \small
    \resizebox{\columnwidth}{!}{
    \begin{tabular}{lc|cccc} 
        \toprule
        \textbf{Method} & 
        \textbf{Dataset} & 
        $\mathbf{Acc}_{\mathit{add}}$ & $\mathbf{F1}_{\mathit{add}}$ & $\mathbf{Acc}_{\mathit{asv}}$ & $\mathbf{F1}_{\mathit{asv}}$ \\
        \midrule
        
        \multirow{2}{*}{Random} & ASVSpoof & 0.500 & 0.500 & 0.500 & 0.500 \\
        & CosyFish & 0.500 & 0.500 & 0.500 & 0.414 \\
        \midrule

        
        \multirow{2}{*}{Qwen2-Audio-Instruct} & ASVSpoof & 0.236 & 0.377 & 0.118 & 0.211  \\
        & CosyFish & 0.023 & 0.000 & 0.151   & 0.262 \\
        \midrule
        \midrule

        \multirow{2}{*}{\textbf{Ours+Zeroshot}} & ASVSpoof & 0.649 & 0.721 &   0.545 & 0.372  \\
        & CosyFish & 0.517 & 0.622  & 0.499 & 0.127  \\
        \midrule

        \multirow{2}{*}{\textbf{Ours+ShortCoT}} & ASVSpoof & \textbf{0.987} & \textbf{0.987}  & 0.520 & 0.152  \\
        & CosyFish & 0.474 & 0.640 & 0.498 & 0.036   \\
        \midrule

        \multirow{2}{*}{\textbf{Ours+CoT}} & ASVSpoof & 0.984 & 0.984 & 0.592 & 0.414  \\
        & CosyFish & 0.500 &  0.6341 & 0.469 & 0.042  \\
        \midrule

        \textbf{Ours+Zeroshot} & ASVSpoof & 0.783 & 0.743 & 0.529 & 0.302 \\
        \textbf{+Unfreeze} & CosyFish & 0.764 & 0.698 & 0.493 & 0.178 \\
        \midrule

        \textbf{Ours+ShortCoT} & ASVSpoof & 0.967 & 0.967 & \textbf{0.751} & \textbf{0.663} \\
        \textbf{+Unfreeze} & CosyFish & 0.483 & 0.641 & \textbf{0.625} & \textbf{0.491} \\
        \midrule
        \midrule
        
        \textbf{Ours+Zeroshot} & ASVSpoof & 0.783 & 0.744 & 0.529 & 0.302 \\
        \textbf{w/ CosyFish} & CosyFish & 0.764 & 0.698 & 0.493 & 0.178 \\
        \midrule
        
        \textbf{Ours+ShortCoT} & ASVSpoof & 0.957 & 0.958 & 0.656 & 0.125 \\
        \textbf{w/ CosyFish} & CosyFish & \textbf{0.951} & \textbf{0.950} & 0.493 & 0.046 \\
        \bottomrule
    \end{tabular}
    }
    \vspace{10pt}
    
    \caption{\textbf{Comparison of our experiment results.} 
    }
    \label{tab:results}
\end{table}
We evaluate our method on two test sets under seven configurations. We first compare three frozen-LLM settings: \textsc{ZeroShot}, \textsc{ShortCoT}, and \textsc{CoT}. We then repeat \textsc{ZeroShot} and \textsc{ShortCoT} with the LLM unfrozen (\textsc{Unfreeze}). Finally, we retrain \textsc{ZeroShot} and \textsc{ShortCoT} on a mixture of the original training data and the synthesized CosyFish data to assess cross-domain generalization.

For the Qwen2-Audio-Instruct baseline, we observe frequent deviations from the required output schema (e.g., malformed outputs or predicting \texttt{unknown}). Following our evaluation protocol, invalid outputs are treated as abstentions, which reduces its measured accuracy below random guessing.

\subsection{Analysis}

\noindent\textbf{Takeaway.} Chain-of-thought (CoT) supervision is the key driver of our gains: it substantially improves evidence-grounded audio deepfake detection (\textsc{ADD}) even with LLM frozen, while \textsc{ASV} remains comparatively harder and benefits more from additional adaptation. 

\noindent\textbf{Effect of CoT supervision.}
Across frozen-LLM settings, chain-of-thought supervision yields a clear and consistent gain on \textsc{ADD}. Moving from \textsc{ZeroShot} to \textsc{CoT} increases \textsc{ADD} accuracy from $0.649$ to $0.984$, and the corresponding \textsc{ShortCoT} variant further improves to $0.987$ (Table~\ref{tab:results}). Importantly, these improvements are achieved without updating the LLM, suggesting that CoT acts as an effective training signal that encourages the projector to align acoustic evidence with the downstream decision. The fact that \textsc{ShortCoT} slightly outperforms full \textsc{CoT} indicates that a concise, structured rationale may reduce prompt/annotation noise and provide a cleaner supervision target than overly verbose reasoning.

\noindent\textbf{Main results and task disparity.}
While CoT substantially strengthens \textsc{ADD}, the corresponding \textsc{ASV} scores remain low when the LLM is frozen (e.g., $\textbf{Acc}_{asv}=0.520$ for \textsc{ShortCoT} on ASVSpoof), suggesting that the model often defaults to a dominant speaker-relation hypothesis (frequently predicting \emph{same speaker}) and fails to robustly exploit speaker-discriminative cues. Unfreezing the LLM markedly improves \textsc{ASV} performance: \textsc{ShortCoT+Unfreeze} reaches $\textbf{Acc}_{asv}=0.751$ and $\textbf{F1}_{asv}=0.663$ on ASVSpoof, and remains strong on CosyFish ($0.625/0.491$). This gap between \textsc{ADD} and \textsc{ASV} indicates that \textsc{ASV} is a more challenging task for our model and likely requires stronger speaker-centric supervision and dataset construction.

\noindent\textbf{Generalization under domain shift.}
Models trained exclusively on ASVSpoof generalize poorly to CosyFish, with \textsc{ADD} accuracy near chance across multiple settings (Table~\ref{tab:results}), highlighting the difficulty of transferring to unseen speakers (VoxCeleb2) and stronger, modern TTS systems. However, incorporating a relatively small amount of CosyFish data substantially improves in-domain performance: \textsc{Ours+ShortCoT w/ CosyFish} achieves $0.951$ \textsc{ADD} accuracy on CosyFish while largely preserving performance on ASVSpoof ($0.957$). This suggests that modest in-domain supervision, when paired with CoT-style evidence grounding, can efficiently bridge the distribution gap.

\noindent\textbf{Baseline behavior.}
Qwen2-Audio-Instruct underperforms due to frequent violations of the required output schema (e.g., malformed outputs or \texttt{unknown} labels). Following our evaluation protocol, such invalid outputs are treated as abstentions, which depresses the reported accuracy below random guessing and reflects limited reliability for structured \textsc{ADD}/\textsc{ASV} judgments in our setting.

\begin{table}[h]
\centering

\resizebox{\columnwidth}{!}{
\begin{tabular}{lc*{7}{c}}
\toprule
\multirow{2}{*}{\textbf{Method}} & \multirow{2}{*}{\textbf{Correct?}} & \multicolumn{7}{c}{\textbf{Attacker (dev)}} \\
\cmidrule{3-9}
& & \textbf{Real} &  \textbf{A01} & \textbf{A02} & \textbf{A03} & \textbf{A04} & \textbf{A05} & \textbf{A06} \\
\midrule
\textbf{Ours + ShortCoT} & \textcolor{green}{\ding{52}} & 502 & 86 & 94 & 89 & 75 & 80 & 77 \\
\textbf{Train + Eval} & \textcolor{red}{\ding{55}} & 6 & 0 & 0 & 1 & 1 & 0 & 5 \\
\cmidrule{2-9}
 & \textbf{Acc(\%)} & 98.8 & {\color{blue} 100} & {\color{blue} 100}  & 98.8 & 98.7 & {\color{blue} 100}  & {\color{red} 93.9} \\
\midrule
\textbf{Ours + ShortCoT} & \textcolor{green}{\ding{52}} & 482 & 85 & 91 & 77 & 72 & 73 & 42 \\
\textbf{Train Only} & \textcolor{red}{\ding{55}} & 26 & 1 & 3 & 13 & 4 & 7 & 40 \\
\cmidrule{2-9}
 & \textbf{Acc(\%)} & 94.9 & {\color{blue} 98.8} & 96.8 & 85.6 & 94.7 & 91.3 & {\color{red} 51.2} \\
\bottomrule
\end{tabular}
}
\caption{Attacker (dev) Performance Metrics (A01-A06)}
\label{tab:attacker_dev_metrics}
\end{table}

\begin{table}[h]
\centering

\resizebox{\columnwidth}{!}{
\begin{tabular}{lc*{7}{c}}
\toprule
\multirow{2}{*}{\textbf{Method}} & \multirow{2}{*}{\textbf{Correct?}} & \multicolumn{7}{c}{\textbf{Attacker (dev)}} \\
\cmidrule{3-9}
& & \textbf{Real} &  \textbf{A07} & \textbf{A08} & \textbf{A09} & \textbf{A10} & \textbf{A11} & \textbf{A12} \\
\midrule
 & \textcolor{green}{\ding{52}} & 996 & 81 & 61 & 52 & 70 & 69 & 64 \\
   & \textcolor{red}{\ding{55}} & 12 & 3 & 6 & 15 & 6 & 5 & 19 \\
\cmidrule{2-9}
\textbf{Ours + ShortCoT} & \textbf{Acc(\%)} & 98.8 & 96.4 & 91.0  & 77.6 & 92.1 & 93.2 & 68.8 \\
\cmidrule{2-9}
\textbf{Train Only} & & \textbf{A13} &  \textbf{A14} & \textbf{A15} & \textbf{A16} & \textbf{A17} & \textbf{A18} & \textbf{A19} \\
 & \textcolor{green}{\ding{52}} & 95 & 76 & 63 & 60 & 32 & 34 & 14 \\
 & \textcolor{red}{\ding{55}} & 11 & 3 & 0 & 13 & 54 & 52 & 50 \\
\cmidrule{2-9}
 & \textbf{Acc(\%)} & 89.6 & 96.2 & {\color{blue} 100} & 82.2 & {\color{red} 37.2} & {\color{red} 39.5} & {\color{red} 21.9} \\
\bottomrule
\end{tabular}
}
\caption{Attacker (eval) Performance Metrics (A07-A19)}
\label{tab:attacker_eval_metrics}
\end{table}

\begin{table}[h]
    \centering
    \tiny
    \begin{tabular}{lc|cccc} 
        \toprule
        \textbf{Method} & 
        \textbf{Dataset} & 
        $\mathbf{Acc}_{\mathit{add}}$ & $\mathbf{F1}_{\mathit{add}}$ & $\mathbf{Acc}_{\mathit{asv}}$ & $\mathbf{F1}_{\mathit{asv}}$ \\
        \midrule
        
        \multirow{2}{*}{\textbf{Ours + Zeroshot}} & ASVSpoof & 0.649 & 0.721 &   0.545 & 0.372  \\
        & CosyFish & 0.517 & 0.622  & 0.499 & 0.127  \\
        \midrule

        \multirow{2}{*}{\textbf{Ours + ShortCoT}} & ASVSpoof & 0.500 &  0.000  & 0.500 & 0.000  \\
        & CosyFish & 0.500 & 0.000 & 0.500 & 0.000   \\

        \bottomrule
    \end{tabular}
    \vspace{10pt}
    
    \caption{Results for ablation study (training without acoustic evidence).}
    \label{tab:results2}
\end{table}


\subsection{Ablation Studies}

In this section, we conduct diagnostic analyses and ablation studies to further validate the effectiveness of our evidence-grounded training. We focus on three questions: (i) whether repurposing the original ASVspoof split (i.e., training with additional spoofing attacks) improves robustness, (ii) how performance varies across individual attack methods, and (iii) whether explicit acoustic evidence is necessary for stable learning.

\noindent\textbf{No-overlap guarantee.}
For all studies in this section, we ensure the train/evaluation sets are \emph{utterance-disjoint}: no audio file is shared between training and evaluation, and consequently no paired example overlaps across splits. Pairs are constructed \emph{within} each split to avoid leakage.

\subsubsection{Training on the original ASVspoof2019 train split}
\label{subsub:asvonly}

In our main experiments, we train on ASVspoof2019's original training split \emph{and} repurpose its original evaluation split as additional training data, while using the original development split for evaluation. ASVspoof2019 contains 19 attack methods in total: A01--A06 appear in the original train/dev protocol, while A07--A19 appear only in the original evaluation split. This repurposing exposes the model to a broader spectrum of spoofing mechanisms during training.

To isolate the effect of this design choice, we train the \textsc{ShortCoT} configuration using \emph{only} the original training split (i.e., attacks A01--A06). We then evaluate this model on both the original development split (A01--A06) and the original evaluation split (A07--A19). In this comparison, the pairing and CoT generation procedure remains identical; only the data available during training differs. This setting directly tests whether the model can generalize from a limited set of attacks (A01--A06) to previously unseen spoofing methods (A07--A19).

\subsubsection{Comparison of attack methods}
\label{subsub:attack}

We further analyze performance at the attack-method level for \textsc{Ours+ShortCoT} under two training regimes: (i) \textbf{Train+Eval} (our main training set, i.e., original train plus original evaluation) and (ii) \textbf{Train-only} (original train only, as in \ref{subsub:asvonly}). For each evaluation instance, we record whether the prediction is correct and the corresponding attacker ID (Axx) for spoofed samples. We aggregate results on a selected subset of evaluation data sorted by attacker and report them in Tables~\ref{tab:attacker_dev_metrics} and~\ref{tab:attacker_eval_metrics}.

\noindent\textbf{Findings.}
On the development attackers (A01--A06), \textbf{Train+Eval} is consistently strong across attack types, while \textbf{Train-only} exhibits noticeably larger degradation on certain attacks, indicating that broader exposure to spoofing mechanisms improves robustness even within the same evaluation family. On the evaluation-only attackers (A07--A19), \textbf{Train-only} serves as a direct cross-attack generalization test. The results are substantially more heterogeneous across attackers, with several VC-style attacks exhibiting the largest drops. Overall, these observations suggest that our approach learns transferable spoofing cues, while generalization to unseen VC attacks remains the most challenging regime.

\subsubsection{Training without acoustic evidence}
\label{subsub:noevidence}

Finally, we test whether the model can succeed without explicit evidence grounding. Concretely, we remove the acoustic-evidence fields from the training text and retrain the same model. The results are shown in Table~\ref{tab:results2}.

\noindent\textbf{Findings.}
Removing acoustic evidence causes training to collapse to a degenerate solution (predicting a single label), resulting in chance-level accuracy and near-zero F1. This indicates that, in our setting, \emph{audio encoding alone is insufficient} to reliably adapt an LLM-based classifier to \textsc{ADD}/\textsc{ASV}. Instead, explicitly surfaced acoustic evidence provides a crucial grounding signal, making CoT-style supervision informative and enabling the model to learn non-trivial decision rules.

\section{Conclusions}

In this work, we propose an ALLM-based pipeline for automatic speaker verification and audio deepfake detection, enhanced by reasoning guided by acoustic evidence. By combining small-scale LLMs and pretrained audio encoders with a custom-trained audio encoding projector, and by enhancing the inference process of the model with acoustic features as evidence, our model outperformed existing ALLMs and attain a high performance on the deepfake detection task. Acoustic evidence and reasoning ensured the explainability of our method. We developed a novel dataset of audios-CoT pairs to support the training of our model. We conducted extensive experiments to explore various factors affecting the model's performance, and compared them to find the best approach to solving the task.
We believe that our research could serve as a basis and important reference for future works related to ALLM and audio deepfake detection.

\bibliographystyle{IEEEtran}
\bibliography{mybib}
\clearpage

\end{document}